\documentclass[12pt]{article}
\usepackage{graphicx}
\title{Study of a narrow $\pi^+ \pi^-$ peak at about 755 MeV/$c^2$ in 
$\bar{p} n \rightarrow 2 \pi^+ 3 \pi^-$ annihilation at rest}
\author{Mario Gaspero\thanks{e-mail: mario.gaspero@roma1.infn.it} \\ 
{\em Dipartimento di Fisica, Sapienza Universit\`a di Roma, and} \\
{\em Istituto Nazionale di Fisica Nucleare, Sezione di Roma 1} \\
{\em Piazzale Aldo Moro 2, I-00185, Rome, Italy}}
\begin{document}
\maketitle
%
%
\begin{abstract}
A narrow peak in the $\pi^+ \pi^-$ mass distribution was seen by the 
Rome-Syracuse Collaboration in $\bar{p} n \rightarrow 2 \pi^+ 3 \pi^-$
annihilation at rest 39 years ago\@.
The reanalysis of this peak finds a mass of $757.4 \pm 2.6$
MeV/$c^2$ and a width slightly narrower than the experimental resolution.
The evidence of the peak is 5.2 standard deviations.
This state is generated in $(12.4 \pm 2.4)\%$ of the 
$\bar{p} n \rightarrow 2 \pi^+ 3 \pi^-$ annihilations at rest.
No spin analysis is possible with the statistics of the experiment. \\
{\em PACS:\/} \ 13.60Le, 13.75.Cs, 14.40.Cs
\end{abstract}
\hspace*{\parindent}
About forty years ago, the Rome-Syracuse Collaboration collected a sample of 
1496 annihilation at rest in deuterium bubble chamber
\begin{equation}
\bar{p} n \rightarrow 2 \pi^+ 3 \pi^- .
\label{pbn5pi}
\end{equation}
This sample was made by $\bar{p} d \rightarrow  p 2 \pi^+ 3 \pi^-$ 
annihilations selected with the cut at 150 MeV/$c^2$ on the 
proton momentum.
665 events were reconstructed at Rome and 831 events at Syracuse.

The $\pi^+ \pi^-$ mass distribution of this sample was published in 1970 with 
a binning of 20 MeV/$c^2$ \cite{RmSy70}\@.
It showed an excess of about 100 combinations in the 
interval $740 < m < 760$ MeV/$c^2$\@.
The structure of this excess is more evident in the distribution with 10 
MeV/$c^2$ shown in Fig.~1\footnote{
The $\pi^+ \pi^-$ distribution with this binning was not previously 
published.}\@.
However, the Collaboration did not claim the observation of a
{\em new resonance\/} because the properties of the annihilation (\ref{pbn5pi})
were not known at that time, and the coincidence of the peak with the $\rho^0$ 
mass suggested that it was generated by the interference of the $\rho^0$ 
meson with another channel.

\begin{figure}[tbh]
\begin{center}
\includegraphics[height=8cm]{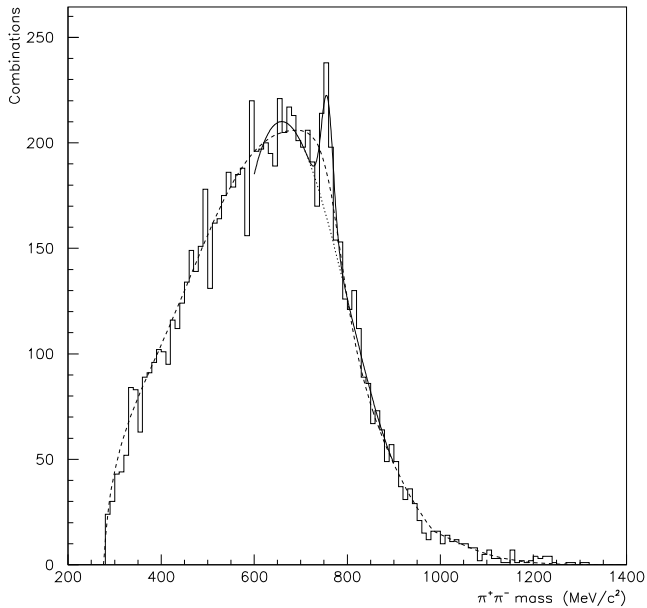}
\caption{$\pi^+\pi^-$ mass ditribution at 10 MeV binning of the 
$\bar{p} n \rightarrow 2 \pi^+ 3 \pi^-$ annihilation at rest.
The dashed curve is the prediction of the fit discussed in Ref.\ 
\cite{Gaspero93}\@.
The solid curve is the result of the fit D\@.
The dotted curve is the third degree polynomial background predicted by 
fit D under the peak.}
\label{figure 1}
\end{center}
\end{figure}

The properties of the annihilation (\ref{pbn5pi}) were understood at the 
beginning of the '90s.
A reanalysis of the same Rome-Syracuse sample \cite{Gaspero92,Gaspero93} 
proved that the annihilation (\ref{pbn5pi}) is dominated by the channel
\[ \bar{p} n \rightarrow f_0(1370) \pi^- , \]
$f_0(1370)$ being a scalar meson \cite{PDG08} that decays into $\rho \rho$ and
$S_w S_w$, where $S_w$ is the symbol for the $\pi^+ \pi^-$ $I = 0$ S-wave 
interaction\footnote{
This interaction was called $\sigma$ in Refs.\ \cite{Gaspero92,Gaspero93}\@.
Now I prefer to use the symbol $S_w$ for it to avoid any confusion with the 
debated resonance $\sigma(600)$ \cite{PDG08}.}.

This reanalysis was able to reproduce satisfactorily ten experimental 
distributions, but failed to reproduce the $\pi^+ \pi^-$ distribution in the 
interval $720 \leq m \leq 770$ MeV/$c^2$, as shown by the dashed curve in 
Fig.~1 that is reporting the prediction plotted in Fig.~3b of Ref.\ 
\cite{Gaspero93}.

A year after the proof of the $f_0(1370)$ dominance in annihilation
(\ref{pbn5pi}) \cite{Gaspero92}, 
the OBELIX Collaboration reported the results of the study of the 
annihilation at low momenta \cite{OBELIX93} 
\begin{equation}
\bar{n} p \rightarrow  3 \pi^+ 2 \pi^- .
\label{nbp5pi} 
\end{equation}
This analysis confirmed the $f_0(1370) \pi^+$ dominan\-ce, followed by the 
$f_0(1370)$ decay into $\rho \rho$ and $S_w S_w$, but did not 
show any peak at about 755 MeV/$c^2$ in the $\pi^+ \pi^-$ distribution.

A structure seen in the annihilation (\ref{pbn5pi}) should also be seen in 
the charge conjugate annihilation (\ref{nbp5pi})\@.
Therefore, at that time, I supposed that the peak shown by Fig.~1 could be 
generated by an exceptional fluctuation higher than 4 standard deviations 
(SD)\@.
But, after the OBELIX paper, Troyan {\em et al.\/} observed 
in $n p \rightarrow n p \pi^+ \pi^-$ interactions at 5.2 GeV/$c$ a narrow 
$\pi^+ \pi^-$ peak at $757 \pm 5$ MeV/$c^2$, with significance 8.5 SD 
\cite{Troyan02}\@.
This result challenged the interpretation of a statistical fluctuation.
For this reason, I have reanalysed the Rome-Syracuse sample.

Table~1 reports two $\chi^2$ fits of the $\pi^+ \pi^-$ mass distribution
shown in Fig.~1, carried out using the MINUIT package \cite{MINUIT} in the 
interval $600 \leq m \leq 900$ MeV/$c^2$\@.
The distribution has been parametrized with the sum of a third degree 
polynomial and 
a Gaussian
\begin{equation}
 D(m) =  P_3( m - m_0 ) + \frac{N_p}{\sqrt{2 \pi} \, \sigma_p} \, 
e^{-\frac{\displaystyle (m-M_p)^2}{\displaystyle 2 \sigma_p^2}} .
\label{distribution}
\end{equation}
The polynomial has been choosen to be a function of the difference 
$m - m_0$ with $m_0 = 750$ MeV/$c^2$ to reduce the correlations between the
coefficients of the polynomial and the number of combinations $N_p$\@.

\begin{table}[tbh]
\begin{center}
\caption{Results of the $\chi^2$ fits of the distribution shown in Fig.~1 
in the interval $600 \leq m \leq 900$ MeV/$c^2$\@.
Fit A is the fit with $\sigma_p$ free.
Fit B is the fit with $\sigma_p$ fixed at the experimental value 12.74
MeV/$c^2$.}
\begin{tabular}{|l|cc|}
\hline
Fit                    &       A         &      B          \\
\hline
$N_p$ (combinations)   & $152 \pm 33$    & $188 \pm 37 $   \\
$M_p$ (MeV/$c^2$)      & $755.4 \pm 1.9$ & $756.3 \pm 2.5$ \\
$\sigma_p$ (MeV/$c^2$) & $7.7 \pm 1.7$   &    12.74        \\
$N_p / \Delta N_p$     &      4.6        &      5.1        \\
$\chi^2$               &     14.7        &     19.6        \\
d.o.f\                 &     24          &      25         \\
$\chi^2$/d.o.f.\       &     0.61        &     0.78        \\
\hline
\end{tabular}
\end{center}
\end{table}

In Table~1 fit A reports the results of the fit of the $\pi^+ \pi^-$ 
distribution shown in Fig.~1 with the resolution $\sigma_p$ free.
Its $\chi^2$ proves that a third degree polynominial is enough for a good fit
in the region of interest.
The signal significance is 4.6 SD, but this fit is unphysical because of its
resolution:
the Rome-Syracuse Collaboration estimated that the experimental resolution of 
the $\pi^+ \pi^-$ masses had a ${\rm FWHM} = 30$ MeV/$c^2$ 
\cite{RmSy70}\footnote{
Actually, Ref.\ \cite{RmSy70} gave the value ${\rm FWHM} = 300$ MeV/$c^2$,
but this was a misprint.}.
Therefore, the physical resolution cannot be lower than $\sigma_p = 12.74$
MeV/$c^2$.

Fit B in Table~1 reports the results of the fit of the same distribution with 
the resolution fixed at 12.74 MeV/$c^2$\@.
This result proves that the peak width is compatible with the experimental 
resolution and improves the significance of the signal to 5.1 SD.

However, the significance of a $\chi^2$ fit may depend on the distribution 
binning.
For this reason, the fits have been repeated using the values of the 
$\pi^+ \pi^-$ masses that were maintained on punched card support\footnote{
These data were already used in an analysis of
the Bose-Einstein correlations \cite{Gaspero95}.}.
The distribution of these masses is in full agreement with that reported in 
Fig.~1\@.
In addition, the study of these data has proved that there are no event 
duplications and that the peak is visible in both the Rome and the
Syracuse subsamples.
This excludes that the peak was generated by a systematic error in the
event reconstruction because the two institutions used two different 
reconstruction program packages.

The knowledge of all the $\pi^+ \pi^-$ masses in the events has also allowed to
repeat the fit with an unbinned log-likelihood 
\[ {\cal L} = - \ln \left[ \prod_n \frac{\displaystyle D(m_n)}{\displaystyle
\int_{600}^{900} D(m) \, dm} \right] , \]
where the product is on the $\pi^+\pi^-$ mass combinations in the 
interval $600 \leq m_n \leq 900$ MeV/$c^2$, and $D(m)$ is the function 
(\ref{distribution})\@.

The results of these fits are reported in Table~2.
The fits C and D in this table have been performed respectively with 
$\sigma_p$ free and with
$\sigma = 12.74$ MeV/$c^2$ and confirm the results of fits A and B\@.
In the following, I will use the results of fit D because it takes into 
account both the known experimental resolution and how the $\pi^+ \pi^-$ 
masses are distributed inside the intervals.

\begin{table}[tbh]
\begin{center}
\caption{Results of the unbinned maximum likelihood fits in the interval 
$600 \leq m \leq 900$ MeV/$c^2$\@.
Fit C is the fit with $\sigma_p$ free.
Fit D is the fit with $\sigma_p = 12.74$ MeV/$c^2$.}
\begin{tabular}{|l|cc|}
\hline
Fit                    &     C            &      D          \\
\hline
$N_p$ (combinations)   & $147 \pm 35 $    & $186 \pm 36$    \\
$M_p$ (MeV/$c^2$)      & $756.7 \pm 1.9$  & $757.4 \pm 2.6$ \\
$\sigma_p$ (MeV/$c^2$) & $ 8.1 \pm 2.1$   &      12.74      \\
$N_p / \Delta N_p$     &      4.2         &     5.2         \\
\hline
\end{tabular}
\end{center}
\end{table}

The study of the correlations between the $\pi^+ \pi^-$ combinations in the 
peak and the other $\pi^+ \pi^-$ masses has shown no evidence for a double 
peak production.
Therefore, the events under the peak are assumed to be generated by the 
channel 
\begin{equation}
\bar{p} n \rightarrow P(757) \pi^+ 2 \pi^- ; \ \  
P(757) \rightarrow \pi^+ \pi^- .
\label{pbnP3pi}
\end{equation}

The punched cards maintained also the $\pi^+ \pi^+$ and $\pi^- \pi^-$ mass
distributions.
The knowledge of all the ten $\pi \pi$ masses has allowed to reconstruct the 
$3 \pi$ and $4 \pi$ masses and other variables.
This analysis has fitted several $\pi^+ \pi^-$ 
distributions obtained by imposing cuts on these variables, but none of them
was able to significantly increase the evidence for the signal.
The conclusion is that the statistics of the Rome-Syracuse sample is too low 
to understand how the signal is correlated with the other three pions.
In particular, no spin analysis is possible.

The results of the fit D allow to draw the following conclusions:
\begin{itemize}
\item The $\pi^+ \pi^-$ mass distribution in the
      $\bar{p} n \rightarrow 2 \pi^+ 3 \pi^-$ annihilation at rest shows a 
      peak at $757.4 \pm 2.6$ MeV/$c^2$ with significance of 5.2 SD.
\item The mass of this peak is in agreement with the value $757 \pm 5$ 
      MeV/$c^2$ of a resonance observed by Troyan {\em et al.\/}\ 
      \cite{Troyan02}.
\item The width of the observed peak is compatible with the experimental
      resolution.
\item The fraction of events with this peak in reaction (\ref{pbn5pi}) is
      \[ f = \frac{N_p}{1496} = (12.4 \pm 2.4)\% . \]
\item The frequency of the $\bar{p} n \rightarrow 2 \pi^+ 3 \pi^-$ annihilation
      at rest is $F_{5 \pi} = (6.9 \pm 0.5)\%$ \cite{Gaspero93}\@.
      Therefore, the frequency of the channel (\ref{pbnP3pi}) is
      \[ F_{P 3\pi} = \frac{N_p}{1496} \, F_{5 \pi} = ( 0.86 \pm 0.18 )\% . \]
\item The spin-parity of this peak cannot be measured using the Rome-Syracuse
      data.
\item An analog peak was not observed in the charge conjugate annihilation 
      (\ref{nbp5pi}) by the O\-BE\-LIX Collaboration \cite{OBELIX93}.
      Therefore, other experimental analyses are needed for confirming or
      disproving the existence of this state.
\end{itemize}
\section*{Acknowledgements}
I would like to thank A. Batosi and F. Batosi who did the boring work of the
manual copying of the $2 \pi$ masses.
I would also like to thank Prof.\ P. Palazzi who has renewed my interest on
this signal.
Lastly, I thank my colleague Dott.\ Fabio Ferrarotto that has helped me to 
revise this paper.

%
%
%

%
\end{document}